\documentclass[english,12pt]{article}
\usepackage{array}
\usepackage{graphicx}
\usepackage{amssymb}
\usepackage{amsmath}
\usepackage{multirow}
\usepackage{prettyref}
\usepackage{babel}
\usepackage{units}
\usepackage[latin1]{inputenc}
\usepackage{amsfonts}
\usepackage{amssymb}
\usepackage{babel}
\usepackage{perpage} 
\MakePerPage{footnote} 
\usepackage{cite}
\def\@fmsl@sh#1#2#3{\m@th\ooalign{$\hfil#1\mkern#2/\hfil$\crcr$#1#3$}}
 \def\eq#1\en{\begin{equation}#1\end{equation}}
\def\s[#1,#2]{[#1\stackrel{\star}{,}#2]}
\def\sx[#1,#2]{[#1\stackrel{\star_{x}}{,}#2]}

\newcommand{\nc}{\newcommand}
\nc{\beq}{\begin{equation}}
\nc{\eeq}{\end{equation}}
\nc{\beqa}{\begin{eqnarray}}
\nc{\eeqa}{\end{eqnarray}}

\def\bc{\begin{center}}
\def\ec{\end{center}}

\def\to{\rightarrow}

\def\gsim{\mathrel{\mathpalette\atversim>}}

\def\bc{\begin{center}}
\def\ec{\end{center}}

\def\gsim{\mathrel{\rlap{\lower4pt\hbox{\hskip1pt$\sim$}}

    \raise1pt\hbox{$>$}}}       

\def\gsim{\mathrel{\rlap{\lower4pt\hbox{\hskip1pt$\sim$}}
    \raise1pt\hbox{$>$}}}       

\begin{document}
\makeatletter
\def\fmslash{\@ifnextchar[{\fmsl@sh}{\fmsl@sh[0mu]}}
\def\fmsl@sh[#1]#2{%
  \mathchoice
    {\@fmsl@sh\displaystyle{#1}{#2}}%
    {\@fmsl@sh\textstyle{#1}{#2}}%
    {\@fmsl@sh\scriptstyle{#1}{#2}}%
    {\@fmsl@sh\scriptscriptstyle{#1}{#2}}}
\def\@fmsl@sh#1#2#3{\m@th\ooalign{$\hfil#1\mkern#2/\hfil$\crcr$#1#3$}}
\makeatother

\thispagestyle{empty}
\begin{titlepage}
\boldmath
\begin{center}
  \Large {\bf Non-locality in Quantum Field Theory due to General Relativity}
    \end{center}
\unboldmath
\vspace{0.2cm}
\begin{center}
{  {\large Xavier Calmet}\footnote{x.calmet@sussex.ac.uk}, {\large Djuna Croon}\footnote{d.croon@sussex.ac.uk} {\large and} {\large Christopher Fritz}\footnote{c.fritz@sussex.ac.uk}}
 \end{center}
\begin{center}
{\sl Physics $\&$ Astronomy, 
University of Sussex,   Falmer, Brighton, BN1 9QH, United Kingdom 
}
\end{center}
\vspace{5cm}
\begin{abstract}
\noindent
We show that General Relativity coupled to a quantum field theory generically leads to non-local effects in the matter sector. These non-local effects can be described by non-local higher dimensional operators which remarkably have an approximate shift symmetry. When applied to inflationary models, our results imply that small non-Gaussianities are a generic feature of models based on General Relativity coupled to matter fields. However, these effects are too small to be observable in the Cosmic Microwave Background.
\end{abstract}  

\end{titlepage}


\newpage

\section{Introduction}

A century after the introduction of General Relativity by Einstein, finding a quantum mechanical description of General Relativity remains one of the holy grails of theoretical physics and one of the few unresolved problems in modern physics. At this stage of our understanding of nature, it is not clear whether the quantization of General Relativity is so difficult because of technical issues, essentially having to deal with a dimensionful coupling constant which is the Planck mass or whether General Relativity or Quantum Mechanics need to be modified at very short distances. Given the current state of the art, it is important to investigate General Relativity and Quantum Mechanics in the energy region where we expect them to work, i.e., below the Planck mass $M_P=1/\sqrt{G_N}$. The concept of effective field theory provides a very powerful framework to investigate the quantization of General Relativity in this energy regime, i.e. below the Planck mass.  Effective field theory methods are a powerful tools to deal with quantum gravity below the Planck mass \cite{Feynman:1963ax,Donoghue:1993eb,Donoghue:1994dn,Donoghue:2012zc,Calmet:2013hfa,Calmet:2014sfa,Calmet:2013hia}. 

An important question is to identify the energy scale at which the effective theory might break down. A well established criterion is that of perturbative unitarity. Treating General Relativity as an effective field theory, several groups have investigated the  gravitational scattering of fields studying whether perturbative unitarity could be violated below the Planck scale \cite{Han:2004wt,Atkins:2010eq,Atkins:2010re,Atkins:2010yg,Antoniadis:2011bi,Aydemir:2012nz,Calmet:2013hia}. It was shown in \cite{Aydemir:2012nz} that in linearized General Relativity with a Minkowski background perturbative unitarity is restored by resumming an infinite series of matter loops on a graviton line in the large $N$ limit, where $N=N_s +3 N_f +12 N_V$ ($N_s$, $N_f$ and $N_V$ are respectively the number of real scalar fields, fermions and spin 1 fields in the model), while keeping $N G_N$ small.  This large $N$ resummation leads to resummed graviton propagator given by
\begin{eqnarray} \label{resprop}
i D^{\alpha \beta,\mu\nu}(q^2)=\frac{i \left (L^{\alpha \mu}L^{\beta \nu}+L^{\alpha \nu}L^{\beta \mu}-L^{\alpha \beta}L^{\mu \nu}\right)}{2q^2\left (1 - \frac{N G_N q^2}{120 \pi} \log \left (-\frac{q^2}{\mu^2} \right) \right)}
\end{eqnarray}
with $L^{\mu\nu}(q)=\eta^{\mu\nu}-q^\mu q^\nu /q^2$, $N=N_s +3 N_f +12 N_V$. A similar calculation has been done by the authors of \cite{Han:2004wt} who have pointed out that the denominator of this resummed propagator has a pair of complex poles which lead to acausal effects (see also \cite{Tomboulis:1977jk,Tomboulis:1980bs} for earlier work in the same direction and where essentially the same conclusion was reached). These acausal effects should become appreciable at energies near $(G_N N)^{-1/2}$. Thus, unitarity is restored but at the price of non-causality.  We shall see that causality can be restored as well by replacing the log term by an interpolating non-local function. However, this procedure does not remove the poles which can be interpreted as black hole precursors \cite{Calmet:2014gya} and correspond to the energy scale at which strong gravitational effects become important and thus the energy scale at which the effective field theory treatment of General Relativity should break down. Note that this scale depends on the number of fields in the theory. Studying quantum effects in General Relativity in the large $N$ limit is not new and has been considered as well by e.g. Smolin \cite{Smolin:1981rm} and Tomboulis \cite{Tomboulis:1977jk,Tomboulis:1980bs}. 

Thought experiments based on General Relativity and quantum mechanics  \cite{Mead:1964zz,Garay:1994en,Padmanabhan:1987au,Calmet:2004mp} lead to the conclusion that distances smaller than the Planck length are not observable. These results can be interpreted as a form of non-locality around the Planck scale. The results obtained in  \cite{Calmet:2014gya} are thus not very surprising. The position of the poles define the energy scale at which the effective theory should break down and the nonlocal effects correspond to the minimal length expected around the mass scale of the first quantum black holes which are extended objects of the size of the inverse of the Planck mass.

The consequences of these non-local effects for the FLRW metric have been investigated in \cite{Donoghue:2014yha}.
The aim of the paper is derive an effective field theory for a scalar field, such as the inflaton, coupled to General Relativity. We will show that this gives rise to some non-local effects in the interactions of this scalar field. These results only assume linearized General Relativity and quantum field theory and are as such non-speculative.

\section{Effective theory and non-locality}

The tree-level gravitational scattering of two scalars has been considered already \cite{Huggins:1987ea}. The invariant amplitude is given by
\begin{eqnarray}
A_{tree}&=&16 \pi G \left ( m^4 \left (\frac{1}{s}+\frac{1}{t}+\frac{1}{u} \right ) +\frac{1}{2s} (2m^2+t )(2m^2+u) \right . \\ \nonumber && \left .
+\frac{1}{2t} (2m^2+s )(2m^2+u)+\frac{1}{2u} (2m^2+s )(2m^2+t) \right )
\end{eqnarray}
with $s=-(p_1+q_1)^2=(p_2+q_2)^2$, $t=-(p_1-p_2)^2=(q_1-q_2)^2$ and $u=-(p_1-q_2)^2=(p_2-q_1)^2$. Note that we are using the signature $(+,-,-,-)$. It is straightforward to calculate the dressed amplitude using the resummed graviton propagator (\ref{resprop}). Let us rewrite
\begin{eqnarray}
i D^{\alpha \beta,\mu\nu}(q^2) = \frac{P^{\alpha \beta,\mu\nu}(q^2)}{1+f(q^2)},
\end{eqnarray}
where $P^{\alpha \beta,\mu\nu}(q^2)$ is the usual graviton propagator and where $f(q^2)$ is given by
\begin{eqnarray}
f(q^2)=-\frac{N G_N q^2}{120 \pi} \log \left (-\frac{q^2}{\mu^2} \right). 
\end{eqnarray}
The dressed amplitude is then given by
\begin{eqnarray}
A_{dressed}&=&16 \pi G \left ( m^4 \left (\frac{1}{s (1+f(s))}+\frac{1}{t(1+f(t))}+ \frac{1}{u(1+f(u))} \right )
 \right . \\ \nonumber && \left .
 +\frac{1}{2s(1+f(s))} (2m^2+t )(2m^2+u)+\frac{1}{2t(1+f(t))} (2m^2+s )(2m^2+u) \right . \\ \nonumber && \left .
+\frac{1}{2u(1+f(u))} (2m^2+s )(2m^2+t) \right ).
\end{eqnarray}
We can now Taylor expand this amplitude  around the massive pole of the dressed propagator and obtain
\begin{eqnarray}
A_{dressed}&=& A_{tree}+ A^{(1)} +\ldots
\end{eqnarray}
with 
\begin{eqnarray}
A^{(1)}&=&\frac{2}{15} G_N^2 N
  \left( m^4 \left (\log \left (-\frac{s t u }{\mu^6} \right) \right )
 \right . \\ \nonumber && \left .
 +\log \left (-\frac{s}{\mu^2} \right) (2m^2+t )(2m^2+u)+\log \left (-\frac{t}{\mu^2} \right) (2m^2+s )(2m^2+u) \right . \\ \nonumber && \left .
+\log \left (-\frac{u}{\mu^2} \right) (2m^2+s )(2m^2+t) \right ).
\end{eqnarray}
It is easy to see that $A^{(1)}$ can be obtained from the following non-local dimension 8 effective operator $O_8$:
\begin{eqnarray}
O_{8}&=&\frac{2}{15} G_N^2 N \left  (  \partial_\mu \phi(x) \partial^\mu \phi(x) - m^2   \phi(x)^2 \right ) \log \left (-\frac{\Box}{\mu^2} \right)  \left( \partial_\nu \phi(x) \partial^\nu \phi(x) - m^2   \phi(x)^2 \right ),
\end{eqnarray}
where $\Box =g^{\mu\nu} \partial_\mu \partial_\nu$.  We emphasize that this calculation is done in linearized General Relativity with a Minkowski background. 

We now need to impose causality on our effective operator. We follow the procedure outlined in \cite{Espriu:2005qn} and \cite{Donoghue:2014yha} to generate a causal action. This requires a reinterpretation of the $\log$-term which can be interpreted as an interpolating non-local function of the type ${\cal L}(x,y)$.  We consider the following action
\begin{eqnarray}
S&=&\int d^4x    \sqrt{-g} \left(  \frac{1}{16 \pi G_N} R(x)- \frac{1}{2} \partial_\mu \phi(x) \partial^\nu \phi(x) +\frac{m^2}{2} \phi^2(x) 
\right . \\ \nonumber && \left .
+\frac{2}{15} G_N^2 N 
  \left (\left  (  \partial_\mu \phi(x) \partial^\mu \phi(x) - m^2   \phi(x)^2 \right ) \log \left (-\frac{\Box}{\mu^2} \right)  \left( \partial_\nu \phi(x) \partial^\nu \phi(x) - m^2   \phi(x)^2 \right) 
  \right) \right ),
\end{eqnarray}
where the $\log$ term is interpreted as an interpolating function
\begin{eqnarray}
S&=&\int d^4x  d^4y  \sqrt{-g} \left(  \frac{1}{16 \pi G_N} R- \frac{1}{2} \partial_\mu \phi(x) \partial^\nu \phi(x) + \frac{m^2}{2} \phi^2 
\right . 
 \\ \nonumber &&   \left .
  +\left (\frac{2}{15} G_N^2 N \right)  \times
  \right .   \\ \nonumber && \left . 
  \left (\left  (  \partial_\mu \phi(x) \partial^\mu \phi(x) + m^2   \phi(x)^2 \right ) \int d^4y \sqrt{-g(y)} \langle x | \log \left (-\frac{\Box}{\mu^2} \right) |y \rangle \left( \partial_\nu \phi(y) \partial^\nu \phi(y) - m^2   \phi(y)^2 \right)  \right ) \right).
\end{eqnarray}
Let us define the interpolating function by
\begin{eqnarray}
{\cal L}(x,y) = \langle x | \log \left (-\frac{\Box}{\mu^2} \right) |y \rangle.
\end{eqnarray}
The specific form for ${\cal L}(x,y)$ depends on the system to which we want to apply this effective theory.  It is straightforward to find a specific representation for a flat Minkowski background. One can use the following approximation valid for small $\epsilon$: $\log(x) \approx -1/\epsilon + x^\epsilon/\epsilon$. One then finds \cite{Espriu:2005qn}:
\begin{eqnarray}
-\langle x |\frac{1}{\epsilon} |y \rangle+ \langle x | \frac{(\Box/\mu^2)^\epsilon}{\epsilon}|y \rangle &=&
-\frac{1}{\epsilon}\delta(x-y) + \frac{1}{\epsilon} \frac{2 \pi^2}{\mu^{2 \epsilon}} \int d^4k k^{2+2\epsilon} \frac{1}{|x-y|} J_1(k|x-y|)
\\ \nonumber &\sim& -\frac{1}{\epsilon}\delta(x-y) - \frac{8 \pi^2}{\mu^{2 \epsilon}} \frac{1}{|x-y|^{4+2\epsilon}}, 
\end{eqnarray}
where $J_1$ is the Bessel function. We see that  that ${\cal  L}(x,y)$ is a function of $x-y$. For a purely time dependent problem in curved space-time (i.e. in cosmology), ${\cal  L}(x,y)$ takes the form
\begin{eqnarray}
{\cal L}(t,t^\prime)=-2 \lim_{\epsilon \to 0} \left (\frac{ \Theta(t-t^\prime-\epsilon)}{t-t^\prime} + \delta(t-t^\prime) (\log(\mu \epsilon)+\gamma \right),
\end{eqnarray}
which is the appropriate form for the in-in formalism \cite{Donoghue:2014yha}. We use this representation in the next section where we study the effects of this effective operator in the CMB. However, remarkably, our calculations do not depend on the specific form of ${\cal L}(x,y)$, we shall merely require that ${\cal L}(0,0)=1$. 

In this section, we have shown that the non-locality induced in the resummed graviton propagator leads to non-locality in the self-interactions of a scalar field coupled to graviton. The same would be true of any spin state as well. Non-locality is an intrinsic feature of a quantum mechanical description of General Relativity as emphasized in the introduction. Note that remarkably, the higher dimensional scalar field operator  obtained by integrating out the poles (quantum black holes) in the graviton propagator,  are invariant under approximative shift symmetry ($\phi \to \phi +c$, where $c$ is a constant) in the limit of the mass of the scalar field going to zero. The breaking of this shift symmetry is proportional to the mass of the scalar field. If we apply this construction to an inflation scenario as we shall do below, this implies that any contribution to the flatness of the potential will be suppressed by powers of $m_s/M_P$ where $m_s$ is the inflaton mass which is of the order of $10^9$ GeV. Quantum gravitational effects arising from quantum black holes are thus small and cannot affect the flatness of the potential. A potential for the scalar field may lead to breaking of the shift symmetry, however one of our main points is that such a symmetry breaking will not be generated by quantum effects in General Relativity if not introduced explicitly in the model. We shall now consider non-local effects due to the dimension 8 operator introduced in this section. We stress that this operator is an intrinsic feature of General Relativity and scalar fields coupled to gravity.

\section{Bounds from Cosmic Microwave Background}

We can now study the implications of this non-local effect which is purely obtained by considering quantum field theory coupled to General Relativity. These effects will be imprinted on the CMB as a deviation in the speed of sound. Focussing on $O_{8}$, we consider the $x$-dependent Lagrangian for a inflaton
\begin{eqnarray}
L(x)&=& X + \frac{m^2}{2} \phi^2(x) + \\ &&  \nonumber +  \frac{8}{15} G_N^2 N \left (X(x)+ \frac{m^2}{2} \phi^2(x) \right) \int d^4y \sqrt{-g(y)}  {\cal L}(x,y)\left (X(y)+ \frac{m^2}{2} \phi^2(y) \right),
 \end{eqnarray}
where we have introduced the standard notation $X(x)=-1/2 \partial_\mu \phi(x) \partial^\mu \phi(x)$ and $X(y)=-1/2 \partial_\mu \phi(y) \partial^\mu \phi(y)$. The speed of sound can be calculated using the standard procedure \cite{Garriga:1999vw} since $L(x)$ is a polynomial in $X$, keeping in mind that $dX(y)/dX(x)=\delta(y-x)$ and that ${\cal L}(0,0)=1$, we find
\begin{eqnarray}
c_s^2=\frac{L(x)_{,X(x)}}{L(x)_{,X(x)}+ 2 X(x) L(x)_{,X(x)X(x)}}\approx 1 - \frac{32}{15} X(x) G_N^2 N 
\end{eqnarray}
which remarkably does not depend on the specific representation chosen for ${\cal L}(x,y)$ to leading order in the $\sqrt{G_N}$ expansion. Restricting ourselves to a spatially homogeneous a scalar field we get
\begin{eqnarray}
c_s \approx 1 - \frac{8}{15} \dot \phi^2 G_N^2 N \approx 1 -\frac{2}{15 \pi}  H^2 \epsilon G_N N,
\end{eqnarray}
where  $H$ is the Hubble parameter, $\epsilon=1/(16 \pi G_N) 1/V^2 (\partial V(\phi)/\partial \phi)^2$ is the slow roll parameter (the slow roll condition is that $\epsilon \ll 1$) and where we have used the approximation $X \approx X L_{,X}$. Quantum effects in General Relativity thus lead to a speed of sound which is not exactly one but close to it. This is a generic feature of General Relativity coupled to matter. Small non-Gaussianities are expected to appear in models of inflation based on General Relativity and quantum field theory even in inflationary models with just one scalar field. However, these effects are too small to be observable since the speed of sound would typically be close to unity.  

Finally, we emphasize that while $O_8$  leads to the leading contribution to deviations in the speed of sound, graviton loop corrections to the scalar propagator may be present, they will be imprinted differently in the Cosmic Microwave Background \cite{Bartolo:2010im}.

\section{Conclusion}
In this paper we have shown that General Relativity coupled to scalar fields naturally leads to non-local effects. This non-locality can be associated with the existence of black hole precursors or quantum black holes \cite{Calmet:2014gya}. We have shown that the amount of non-locality is determined by the number of matter fields in the theory since it determines the location of the poles in the resummed graviton propagator. General Relativity induces non-local effects in the scalar field sector. These effects can be described in terms of an effective higher non-local dimensional operator which remarkably has an approximate shift symmetry. When applied to inflationary models, we have shown that these non-local effects lead to a small  non-Gaussianities in models of inflation involving a scalar field and General Relativity. 

{\it Acknowledgments:}
We would like to thank Veronica Sanz for helpful discussions. This work is supported in part  by the Science and Technology Facilities Council (grant number  ST/J000477/1).


\bigskip{}


\baselineskip=1.6pt 

\begin{thebibliography}{10}

\bibitem{Feynman:1963ax} 
  R.~P.~Feynman,
  Acta Phys.\ Polon.\  {\bf 24}, 697 (1963).
\bibitem{Donoghue:1993eb} 
  J.~F.~Donoghue,
  Phys.\ Rev.\ Lett.\  {\bf 72}, 2996 (1994)
  [gr-qc/9310024].
\bibitem{Donoghue:1994dn} 
  J.~F.~Donoghue,
  Phys.\ Rev.\ D {\bf 50}, 3874 (1994)
  [gr-qc/9405057].
\bibitem{Donoghue:2012zc} 
  J.~F.~Donoghue,
  AIP Conf.\ Proc.\  {\bf 1483}, 73 (2012)
  [arXiv:1209.3511 [gr-qc]].
\bibitem{Calmet:2013hfa} 
  X.~Calmet,
  Int.\ J.\ Mod.\ Phys.\ D {\bf 22}, 1342014 (2013)
  [arXiv:1308.6155 [gr-qc]].
\bibitem{Calmet:2014sfa} 
  X.~Calmet,
  arXiv:1412.6758 [gr-qc].
 
  
\bibitem{Calmet:2013hia} 
  X.~Calmet and R.~Casadio,
  Phys.\ Lett.\ B {\bf 734}, 17 (2014)
  [arXiv:1310.7410 [hep-ph]].
  
\bibitem{Han:2004wt} 
  T.~Han and S.~Willenbrock,
  Phys.\ Lett.\ B {\bf 616}, 215 (2005)
  [hep-ph/0404182].
  
\bibitem{Aydemir:2012nz} 
  U.~Aydemir, M.~M.~Anber and J.~F.~Donoghue,
  Phys.\ Rev.\ D {\bf 86}, 014025 (2012)
  [arXiv:1203.5153 [hep-ph]].
  
  
\bibitem{Atkins:2010eq} 
  M.~Atkins and X.~Calmet,
  Phys.\ Lett.\ B {\bf 695}, 298 (2011)
  [arXiv:1002.0003 [hep-th]].

\bibitem{Atkins:2010re} 
  M.~Atkins and X.~Calmet,
  Eur.\ Phys.\ J.\ C {\bf 70}, 381 (2010)
  [arXiv:1005.1075 [hep-ph]].
  
\bibitem{Atkins:2010yg} 
  M.~Atkins and X.~Calmet,
  Phys.\ Lett.\ B {\bf 697}, 37 (2011)
  [arXiv:1011.4179 [hep-ph]].
    
\bibitem{Antoniadis:2011bi} 
  I.~Antoniadis, M.~Atkins and X.~Calmet,
  JHEP {\bf 1111}, 039 (2011)
  [arXiv:1109.1160 [hep-ph]].
  



  




\bibitem{Tomboulis:1977jk} 
  E.~Tomboulis,
  Phys.\ Lett.\ B {\bf 70}, 361 (1977).
  

\bibitem{Tomboulis:1980bs} 
  E.~Tomboulis,
  Phys.\ Lett.\ B {\bf 97}, 77 (1980).



  
\bibitem{Calmet:2014gya} 
  X.~Calmet,
  Mod.\ Phys.\ Lett.\ A {\bf 29}, 1450204 (2014)
  [arXiv:1410.2807 [hep-th]].
  
\bibitem{Smolin:1981rm} 
  L.~Smolin,
  Nucl.\ Phys.\ B {\bf 208}, 439 (1982).

\bibitem{Mead:1964zz} 
  C.~A.~Mead,
  Phys.\ Rev.\  {\bf 135}, B849 (1964).
\bibitem{Garay:1994en} 
  L.~J.~Garay,
  Int.\ J.\ Mod.\ Phys.\ A {\bf 10}, 145 (1995)
  [gr-qc/9403008].
\bibitem{Padmanabhan:1987au} 
  T.~Padmanabhan,
  Class.\ Quant.\ Grav.\  {\bf 4}, L107 (1987).
\bibitem{Calmet:2004mp} 
  X.~Calmet, M.~Graesser and S.~D.~H.~Hsu,
  Phys.\ Rev.\ Lett.\  {\bf 93}, 211101 (2004)
  [hep-th/0405033].
  
  
  
\bibitem{Donoghue:2014yha} 
  J.~F.~Donoghue and B.~K.~El-Menoufi,
  Phys.\ Rev.\ D {\bf 89}, 104062 (2014)
  [arXiv:1402.3252 [gr-qc]].
  

\bibitem{Huggins:1987ea} 
  S.~R.~Huggins and D.~J.~Toms,
  Class.\ Quant.\ Grav.\  {\bf 4}, 1509 (1987).
   
\bibitem{Anber:2010uj} 
  M.~M.~Anber, J.~F.~Donoghue and M.~El-Houssieny,
  Phys.\ Rev.\ D {\bf 83}, 124003 (2011)
  [arXiv:1011.3229 [hep-th]].
  
\bibitem{Espriu:2005qn} 
  D.~Espriu, T.~Multamaki and E.~C.~Vagenas,
  Phys.\ Lett.\ B {\bf 628}, 197 (2005)
  [gr-qc/0503033].
    
\bibitem{Garriga:1999vw} 
  J.~Garriga and V.~F.~Mukhanov,
  Phys.\ Lett.\ B {\bf 458}, 219 (1999)
  [hep-th/9904176].
    
\bibitem{Bartolo:2010im} 
  N.~Bartolo, M.~Fasiello, S.~Matarrese and A.~Riotto,
  JCAP {\bf 1012}, 026 (2010)
  [arXiv:1010.3993 [astro-ph.CO]].
\end{thebibliography}
\end{document}